\documentclass[prb,twocolumn,amsmath,amssymb,showpacs,letterpaper]{revtex4-1}
\usepackage{graphicx}
\usepackage{verbatim}
\begin{document}
\title{A Hybrid Analog/Digital Phase-Locked Loop for Frequency Mode Non-contact Scanning Probe Microscopy}
\author{M. M. Mehta}
\author{V. Chandrasekhar}
\email{v-chandrasekhar@northwestern.edu}
\affiliation{Department of Physics and Astronomy, Northwestern University, Evanston, Illinois. 60208, USA}



\begin{abstract}
Non-contact scanning probe microscopy (SPM) has developed into a powerful technique to image many different properties of samples.  The conventional method involves monitoring the amplitude, phase or frequency of a cantilever oscillating at or near its resonant frequency as it is scanned across the surface of a sample.  For high $Q$  factor cantilevers, monitoring the resonant frequency is the preferred method in order to obtain reasonable scan times.  This can be done by using a phase-locked-loop (PLL).  PLLs can be obtained as commercial integrated circuits, but these do not have the frequency resolution required for SPM.  To increase the resolution, all-digital PLLs  requiring sophisticated digital signal processors or field programmable gate arrays have also been implemented.  We describe here a hybrid analog/digital PLL where most of the components are implemented using discrete analog integrated circuits, but the frequency resolution is provided by a direct digital synthesis chip controlled by a simple 
PIC microcontroller.  The PLL has excellent frequency resolution and noise, and can be controlled and read by a computer via a USB connection.     
\end{abstract}

\maketitle
Non-contact scanning probe microscopy (SPM) in its various forms is a powerful tool to image and study the properties of samples near their surfaces.\cite{kalinin}  In its simplest form for atomic force microscopy (AFM), a cantilever with a tip is driven at its mechanical resonance frequency $f_0$ near the surface of the sample.  As the tip is brought very close to the surface, the interaction of the tip with the sample modifies the mechanical resonance frequency.  If the driving frequency is kept constant, the amplitude and phase of the resulting cantilever oscillation will change.  Both the amplitude and phase of the oscillation can be used as feedback signals to maintain the tip at a fixed distance as it is scanned across the surface.  Amplitude or phase mode AFM has been used very successfully to obtain topographical and other scanning probe images.

Amplitude mode AFM suffers from one serious drawback.  If the quality factor, $Q$, of the cantilever is large, it takes a time on the order of $\tau_r \sim Q/f_0$ for the amplitude to relax to its steady-state value when the cantilever tip is moved.  In vacuum and/or at low temperatures, the $Q$ of a cantilever can approach $10^5-10^6$, resulting in extraordinarily long times to obtain even a low resolution image.  Phase mode scanning does not suffer from this problem, as the phase relaxes almost instantaneously (on a time scale of order $\tau_0 \sim 1/f_0$) to its steady state value.  However, in order to use phase mode imaging, the shift in frequency of the cantilever on approaching the surface must be well within the envelope of the resonance curve, determined by the halfwidth of the resonance $\Delta f \sim f_0/Q$.  If the $Q$ is large, this may not always be the case, and one is then restricted to imaging at distances relatively far from the surface.

Tracking the change in resonant frequency directly avoids many of these problems.\cite{giessibl}  As with phase mode imaging, the frequency changes almost instantaneously on time scales of the order of $\tau_0$, but since one is always on resonance, changes in frequency many times $\Delta f$ can be tracked.  While tracking the resonant frequency can be done using frequency modulation techniques, more recently, phase-locked-loops (PLLs) are being employed.  PLLs have gained in popularity in the last few decades,\cite{horowitz} and are now used extensively in digital electronics to generate timing control signals that are precisely aligned to a master clock.
\begin{figure}
\center{\includegraphics[width=9cm]{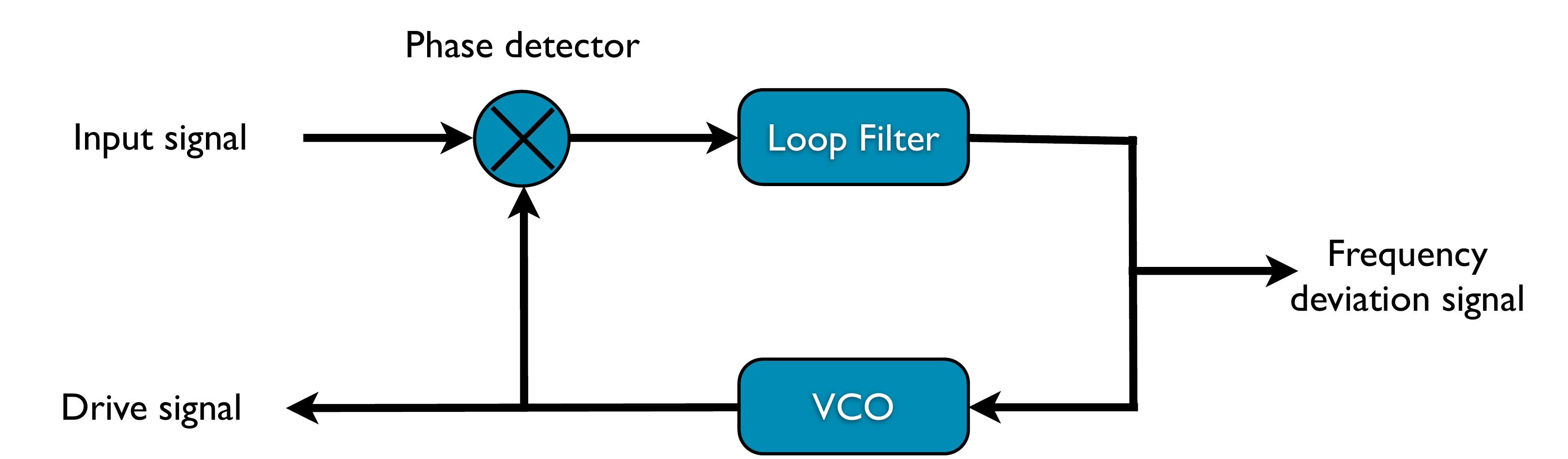}}
\caption{\small{Schematic of a phase-locked loop.}}
\label{PLLSchematic}
\end{figure}

Figure 1 shows a schematic of a PLL.\cite{horowitz}  The signal to be tracked is one input to a phase detector (PD), the second input of which comes from a local oscillator in the form of a voltage controlled oscillator (VCO).  In the simplest case, for analog signals, which is what we are concerned with here, the PD is a four-quadrant analog multiplier.  Let $A_s \cos (2 \pi f_s t + \phi)$ be the input signal, and $A_{VCO} \cos (2 \pi f_{VCO} t)$ be the signal from the VCO.  ($\phi$ represents any fixed phase difference between the two signals.)  Assume initially that the frequencies $f_s$ and $f_{VCO}$ are close to each other.  The output of the phase detector then has components at the sum and difference frequencies
\begin{equation*}
\sim \cos(2 \pi (f_s - f_{VCO})t + \phi) + \cos(2 \pi (f_s + f_{VCO}) t + \phi).
\end{equation*}                 
We are interested here in the component at the difference frequency.  To eliminate the component at the sum frequency, the output of the PD is fed to a filter (the loop filter).  The simplest implementation of the loop filter is just a passive low-pass filter with a cut-off frequency well below the sum frequency.  The output of the loop filter is fed to the input of the VCO.  If the loop filter and VCO parameters are correct, the VCO will quickly ``lock'' in to the input signal, matching its frequency and maintaining a fixed relative phase.  At this point, the output of the loop filter is a dc voltage that is a measure of the deviation of the frequency being generated by the VCO from its free running or center frequency.  If the center frequency of the VCO is already close to the input frequency, the output of the loop filter will be close to zero.  In this case, the phase difference $\phi           \sim \pi/2$, and the input signal and signal generated are almost in quadrature.

The PLL circuit shown in Fig. \ref{PLLSchematic} appears ideal for driving a non-contact mode transducer of a scanning probe microscope, which is usually driven at or near its resonance frequency. The input to the PLL is the amplified signal from the force transducer of the microscope, typically a cantilever, or as is the case in our own instrument, a tuning fork transducer; the output of the VCO also drives the transducer.  The center frequency of the VCO is set when the tip of the probe is far from the surface.  The frequency deviation signal can then be used as the feedback signal for the SPM control software.  For example, for an atomic force microscope (AFM), a fixed frequency deviation would correspond to a fixed distance of the tip from the surface. 

\section{Discrete PLL based on lock-in amplifier and waveform generator}

PLLs covering a wide range of frequencies are available commercially as integrated circuits (ICs).  However, for the audio range of interest here (typically 10-100 kHz), the frequency resolution of these ICs is not sufficient for SPM applications.\cite{ti}  For example, for the tuning fork transducers that we use in our microscopes, which have resonant frequencies of around 32 KHz, we would like to be able to detect changes in frequency of the order of 1 mHz.   In addition, the center frequency of the PLL and the frequency sensitivity (Hz per volt of input) are fixed by discrete components and not easy to change.   For these reasons, it is convenient to replace the analog VCO by what is called a numerically controlled oscillator (NCO), or Direct Digital Synthesis (DDS) ICs.\cite{dds}  The operation of these chips is based on phase accumulators:  the value of the phase is incremented by a specific value on every cycle of the master clock.  This value of the phase is then used to generate an analog output by 
using a sine wave lookup table.  By modifying the phase increment, one can modify the frequency of the output sine wave.  Since the digital width of the accumulator can be 32 bits, one can obtain a frequency resolution of 1 part in 2$^{32}$ of the master clock frequency, or better than 1 part in 4$\times$10$^9$.  The frequency sensitivity can also be set programmatically to any desired value.

Unfortunately, to control such a NCO or DDS chip to provide a replacement for the VCO, one requires a computer.  This computer would need to read the output of the loop filter using an analog-to-digital converter (ADC), calculate the required frequency change, and appropriately program the chip.  We shall describe later exactly this implementation, but first we shall describe how one can assemble a PLL with the required frequency resolution with common laboratory instruments, specifically a lock-in amplifier (LIA) with an external reference mode, and a frequency-modulated waveform generator.  

At the heart of every LIA is a phase comparator.  To use the LIA as a phase detector, we use the output of the VCO as the reference input of the LIA in external reference mode, and the signal from the SPM transducer to the LIA input.  The in-phase or `$X$' output of the LIA is then the required output of the phase detector.  In most cases, the LIA provides a low pass filter on this output whose time constant can be adjusted, so that in essence the LIA also incorporates the loop filter in Fig. \ref{PLLSchematic}.  
The $X$ output of the LIA is used to drive the FM modulation input of a digital waveform generator, whose center frequency is set to match the resonance frequency of the transducer when the tip is far from the surface.  The sensitivity of the FM modulation input can be usually adjusted over a very wide range, which affects the lock range of the PLL as well as its frequency resolution.  To complete the loop, the output of the waveform generator is used as the reference input of the LIA, and also to drive the transducer.  The $X$ output of the LIA is the frequency deviation signal that can be monitored by the SPM control program.

We have implemented this discrete PLL using an EG\&G 7260 digital LIA\cite{signalrecovery} and an Agilent 33500B waveform generator,\cite{agilent} although almost any digital or analog LIA and any waveform generator with FM modulation capability could be used.  Using a LIA offers some convenient advantages.  First, with more recent digital LIAs, one can simultaneously measure additional quantities such as the amplitude and phase of the signal in addition to the $X$ output.  Second, although theoretically the phase shift between the driving signal and the response of the sensor oscillator is $\pi/2$, in practice, the phase shift as seen at the instrumentation may be different, since the wiring and connectors for the SPM will result in an additional contribution.    With a LIA, one can shift the relative phase between the drive and response by simply adjusting the phase controls, or by adding an offset to the $X$ output.  Since it is not simple \textit{a priori} to determine the additional phase shift 
introduced by the wiring, the protocol we use is to adjust both the phase controls on the LIA and the center frequency of the waveform generator to obtain a maximum in the amplitude of the sensor signal when the SPM tip is far from the surface.  Finally, we note that even the waveform generator is not required if the LIA contains an oscillator module, since the LIA itself incorporates a PLL.  On many LIAs, the signal from the transducer, if it is sufficiently amplified, can be fed into the external reference source; the output of the oscillator then is phase locked to the input, and can be used to drive the transducer.  The problem with this arrangement lies in obtaining a signal that corresponds to the frequency deviation:  most LIAs that we are familiar with do not have an output corresponding to the frequency that has sufficient resolution for our requirements, and reading the frequency shift over a serial, parallel or GPIB connection is too slow.

The frequency resolution of this discrete PLL is determined by the output noise of the LIA and the FM modulation range of the waveform generator.  Ideally, the latter should be as small as possible to increase the resolution, but still have sufficient range to accommodate largest frequency deviations expected.  We typically find that a frequency range of 10-50 Hz is sufficient.  On the Agilent 33500B waveform generator, a voltage of $\pm$5 V corresponds to the full range.  Assuming a full range of 10 Hz, and an output noise of 1 mV rms from the $X$ output of the EG\&G 7260 LIA, we obtain a frequency resolution of about 1 mHz, more than sufficient taking into account that the overall frequency shifts for our tuning fork transducers are of the order of 1 Hz.

There is one major problem with using this LIA-waveform generator PLL that might be specific to our instrumentation, but we feel is worth mentioning.  The time response of the PLL is set by the time constant of the loop filter, which in this case is a low pass filter.  On the EG\&G 7260, the minimum time constant for the $X$ output is 5 ms, corresponding to an update rate of 200 Hz, which sets the minimum time required for the PLL to respond to changes in the resonant frequency of the transducer, and consequently also determines the minimum time to obtain an image.  Ideally, we would like to have a response time an order of magnitude smaller.  One can potentially use other LIAs, but they frequently have other limitations.  For example, we have also used an older PAR 124A analog LIA.\cite{par}  This has a minimum output time constant of 1 ms, but has the tendency to go out of reference lock with even small changes in frequency, and then takes a very long time to relock.  Consequently, although we have been 
able to obtain close-approach curves with the LIA-waveform generator PLL, we have not been able to obtain satisfactory images in scanning.

A more serious concern with the design of the PLL as described above to drive a non-contact transducer in a SPM is associated with the low pass loop filter.  To illustrate the problem, we consider the specific case of a tuning fork transducer.  The tuning fork can be modeled as a forced damped harmonic oscillator whose motion is described by the equation\cite{kleppner}
\begin{equation}
\ddot{x} + \gamma \dot{x} + \omega^2_0 x = \frac{F_0}{m} \cos \omega t
\label{eqn1}
\end{equation}  
Here $\gamma$ is a constant that describes any dissipative forces, $\omega_0$ is the resonant frequency of the oscillator, $m$ is its mass, and $F_0 \cos \omega t$ is the driving force at frequency $\omega$.  (In the lightly damped case, the quality factor is related to $\gamma$ by $Q= \omega_0/\gamma$.)  The steady-state solution of this equation is $x=A \cos(\omega t + \phi)$, where the amplitude $A$ and phase $\phi$ are given by\cite{kleppner}
\begin{align}
A &= \frac{F_0}{m} \frac{1}{[(\omega^2_0 -\omega^2)^2 + (\omega \gamma)^2]^{1/2}}, \\
\phi &= \arctan{\left(\frac{\gamma \omega}{\omega^2 - \omega^2_0}\right)}
\end{align} 
On resonance ($\omega=\omega_0$), the phase shift between the response and the driving force is $\phi=\pi/2$.  For a tuning fork transducer, the forcing term is provided by electrically driving the tuning fork with the VCO signal, and the response is the amplified current through the tuning fork which serves as the input to the PLL, neglecting for the moment any other phase changes that may arise due to phase shifts in connecting wires, for example.

\begin{figure}[t!]
\center{\includegraphics[width=8.5cm]{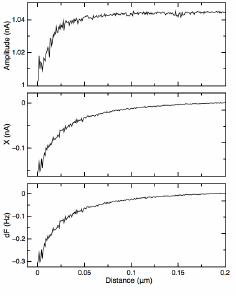}}
\caption{\small{Amplitude, $X$ output and frequency shift measured using the discrete LIA-waveform generator PLL, as described in the text.}}
\label{PseudoPLL}
\end{figure}

Consider now the case when the loop filter of the PLL is a simple low pass filter, and suppose we fix the center frequency of the VCO at the resonant frequency of the tuning fork when the tip of the force sensor is far from the surface of the sample, as described earlier.  In this situation, the two inputs to the phase detector are in quadrature, and the output of the loop filter vanishes.  As the tip is moved closer to the sample, the force interactions between the tip and the sample will modify the resonant frequency of the tuning fork.  The PLL will track this frequency change.  In order to do this, however, the input to the VCO must be different from zero.  This means that the output of the loop filter is now finite, which in turn means that the two inputs to the phase detector are no longer in quadrature, i.e., the tuning fork is not being driven exactly at resonance, since the driving force and response are no longer $\pi/2$ out of phase.  Thus, while the frequency output of the PLL does change as  the 
tip approaches the surface, the tuning fork is being driven progressively off resonance.  This problem is particularly acute for high $Q$ transducers, as in this case,  the phase changes rapidly from 0 to $-\pi$ as $\omega$ is swept through $\omega_0$.  
The second problem with using a simple low pass filter as the loop filter is that its output depends not only on the phase difference between its two inputs, but also on their amplitudes.  One can assume that the VCO output is constant in amplitude, but the amplitude of oscillation of the transducer does change as the tip approaches the surface.  One way to counter this is to have a feedback loop that maintains constant the amplitude of oscillation of the cantilever, but if this is not implemented, the frequency deviation will be a complicated function of the phase change and amplitude change.

To demonstrate these two problems, we show in Fig. \ref{PseudoPLL} a close approach curve taken with the discrete LIA-waveform generator PLL.  This close-approach curve was taken on our home-built tuning-fork based SPM using our real-time scanning probe software RTSPM\cite{chandra} with a commercial AFM tip mounted on a tuning fork transducer.\cite{rozhok}  In addition to the amplitude of the signal from the tuning fork, the plot also shows the $X$ deviation and the frequency shift of the tuning fork.  (Since the $X$ output drives the frequency shift in this configuration, both curves track each other.)  Both the amplitude and the $X$ output (and hence the phase) change as a function of the distance of the tip from the surface, and hence the measured frequency deviation is not a true indication of the actual frequency shift of the tuning fork.

Both these problems can be eliminated by ensuring that the output of the phase detector is always zero when the PLL is locked, for in that case, the two input signals to the phase detector are 90$^\circ$ out of phase, and the oscillation amplitude of the transducer is irrelevant since the output is zero.  To do this, one inserts a proportional-integral (PI) controller after the low pass filter.  The PI controller will try and maintain its input error signal zero, which in this case is the output of the low pass filter.  In lock, the error input to the PI controller vanishes, but because of the integral component, its output is a finite value that is fed to the VCO.  Consequently, one can maintain the output of the phase detector zero, i.e, have its two inputs be in quadrature, but still have a finite frequency deviation.  For our tuning fork transducer, this means that one can track its changes in frequency, but still drive it always on resonance.   
\section{Hybrid analog/digital PLL}
\begin{figure*}[!]
\vspace{0.25cm}
\center{\includegraphics[width=16cm]{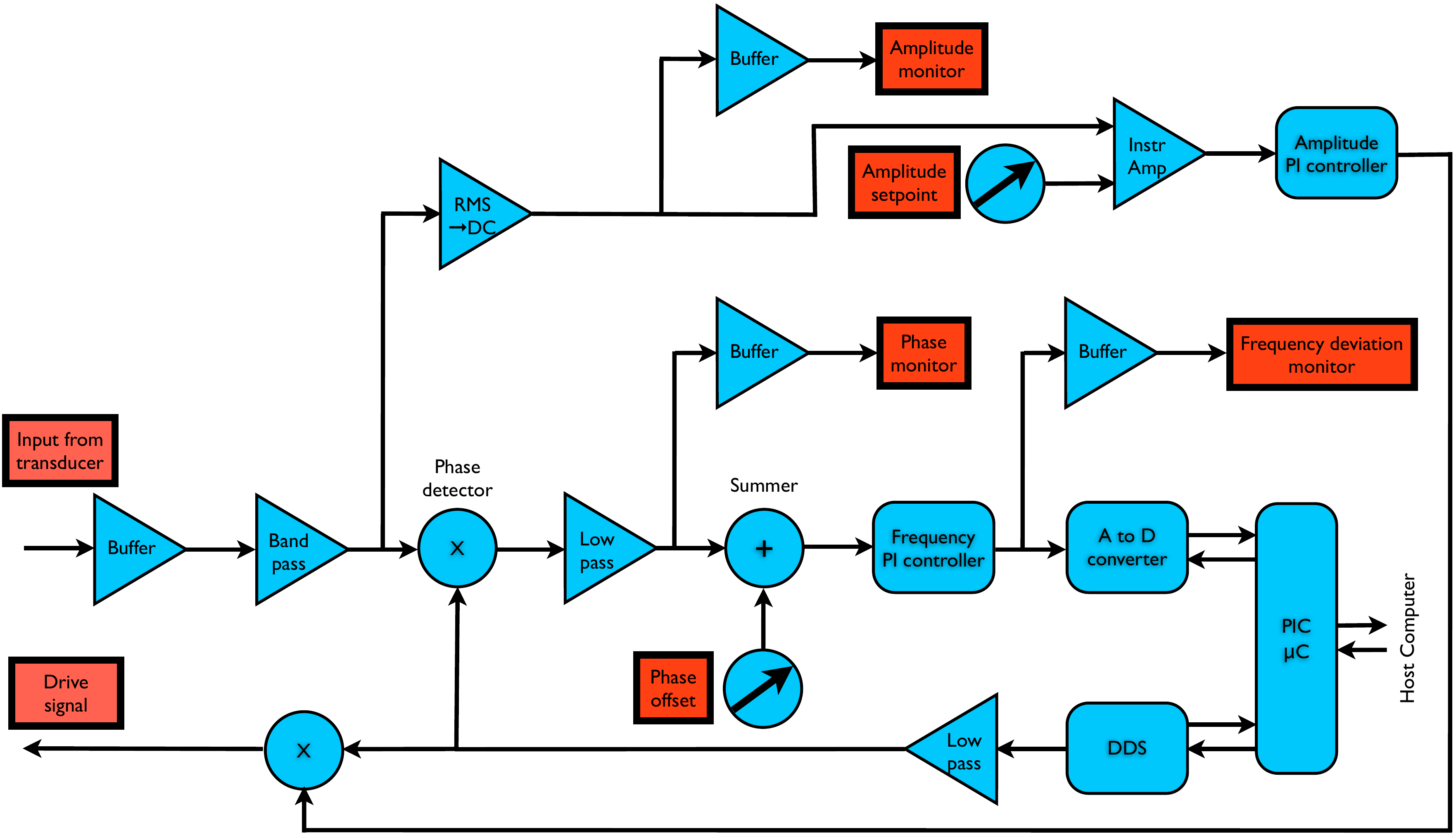}}
\caption{Schematic of the hybrid analog/digital PLL.}
\vspace{-0.25cm}
\label{FullPLL}
\end{figure*}

It is a simple matter to insert a PI controller after the $X$ output of the LIA.  In doing so, it is important to remember that the overall response time is then determined by both the low pass filter of the LIA and the integral time constant of the PI controller.  As we mentioned above, the output time constant of our LIA is quite long.  This, coupled with the fact that we prefer not to dedicate two versatile and generally useful laboratory instruments (the LIA and the waveform generator) exclusively to our SPM led us to develop a hybrid analog/digital PLL.  This PLL incorporates the elements discussed above in the form of discrete circuits, with the VCO being replaced by an ADC, a DDS chip and a microcontroller.

One might reasonably ask that if we are going to use a microcontroller to generate the drive signal, why not go with a fully digital PLL?  A fully digital PLL digitizes the input signal from the transducer using an ADC, but then implements the phase detector and loop filter in software, again using a NCO or DDS chip for output.  The reason is that the computing demands on the microprocessor to implement a digital PLL are not trivial.  To illustrate, for a 32 kHz signal such as that from our tuning forks, one would require a digitizing time step of at least 1 $\mu$s, or a sampling frequency of 1 MHz.  This is just the input sampling requirement:  once one data point is taken, the phase detector and loop filter algorithms must be performed, the frequency deviation must be calculated and the DDS chip programmed, any requests from the host computer must be serviced, and the frequency deviation must either be offered as an analog output or transferred to the host computer, all before the next data point is 
processed.  Consequently, such digital PLLs all use powerful digital signal processors to perform these tasks.\cite{nanosurf}  We have chosen instead to implement most of the PLL as discrete electronic components, using a relatively low power microprocessor (a Microchip PIC controller)\cite{microchip} only to control the DDS chip and to communicate with the host computer.

\subsection{Analog electronics}

Figure \ref{FullPLL} shows a schematic of our PLL design, which incorporates some additional components specifically to interface with SPM control software and a tuning fork transducer.  We describe here the overall design of the instrument; the detailed electronic schematic can be found in the Supplemental Materials\cite{supplemental}. 

The input to the PLL is the amplified response of the transducer, and is typically an ac signal at the level of a volt.  For example, for a tuning fork transducer, one measures the ac current, which is amplified by a current preamplifier.  For our system, typical currents from the tuning fork are in the nA range, corresponding to drive excitations of the order of millivolts.  The current from the tuning fork is amplified using a homemade current preamp (not shown) with a gain $\sim$10$^9$ V/A, so that the input to the PLL is an ac signal of amplitude $\sim$1 V.  After buffering, this input signal is passed through a bandpass filter in multiple feedback architecture with a center frequency of 30 kHz, but with a broad bandpass of about 10 kHz.  The filter is needed to reduce extraneous noise signals, in particular, line frequency noise, which introduces phase jitter in the PLL.  The pass band of the filter is kept broad because the frequency of our tuning forks may vary from 25-34 kHz, depending on the mass of 
the tip and 
glue on the tines of the tuning fork.  The buffer and low pass filters are constructed from common op-amp chips such as the LF356.\cite{ti}

After the bandpass filter, the signal is split up into two paths.  One path goes to a RMS-to-DC converter (Analog Devices AD637).\cite{analog}  The output of this chip gives a measure of the amplitude of the input signal.  After passing through a buffer amplifier, it is connected to a front panel BNC so it can be monitored externally.  For scanning using frequency feedback, it is sometimes desirable to keep the amplitude of oscillation of the tuning fork transducer constant.  The output of the AD637 is also fed as one of the inputs to an amplitude control circuit.  The second input is a dc voltage controlled by a 10-turn potentiometer connected to a voltage reference that can be adjusted from the front panel by the user to set the amplitude of current.  These two signals, the output of the AD637 and the amplitude set point voltage are respectively connected to the positive and negative inputs of an instrumentation amplifier (Analog Devices AD524\cite{analog}) to generate an error signal.  The error signal is 
fed to the input of a PI controller implemented from 4 LF356 op-amps (see electronic schematic).  We use 4 op-amps for the PI controller so that we can set the time constant of the integral component independently of the gain of the proportional component.  These gains are set using trimpots on the circuit board, since once these gains are determined for a particular system, they should not have to be changed repeatedly.  The time constant of the integral component in our circuit is of the order of milliseconds:  since the oscillation amplitude frequently takes longer to achieve its equilibrium value for high $Q$ transducers, there is no need for a faster response.

The output of the bandpass filter is also fed to one input of a phase detector, which in our case is simply a high-speed, four quadrant multiplier (Analog Devices AD632\cite{analog}).  The other input to the multiplier chip is the filtered output of the DDS chip, which we shall discuss below.  The output of the AD632 is passed through a low pass filter to eliminate the component at the sum frequency, as discussed above.  For our tuning forks, this is of the order of 50-65 kHz, so that  a low pass filter with a cutoff of 40 kHz should work.  In reality, we keep the cut-off frequency of the low pass filter at 5 kHz so as to reduce noise, as the output of the low pass filter is connected to a front panel BNC  through a buffer amplifier to enable monitoring of the phase.  

The output of the low pass filter is summed with a dc voltage and then fed to the input of another PI controller identical in design to the amplitude PI controller.  The input dc  voltage is controlled by a 10-turn pot in much the same way as the amplitude set point.  The frequency PI controller tries to maintain the error signal at its input at zero.  In the absence of the dc offset voltage, this means that the output of the low pass filter is zero, or in other words, that the two signals coming in to the phase detector are 90$^\circ$ out of phase.  With a finite dc offset voltage, the PI tries to make the sum of the dc voltage and the low pass filter vanish, which means the phase shift between the two input signals can be modified even when the PLL is in lock.  In practice, we find that the phase can be modified over a range of $\pm 90^\circ$ with very fine control using the 10-turn phase control potentiometer. 

The integral time constant of the frequency PI controller is set to be 0.5 ms as trade-off between a low noise output and a sufficiently rapid response, as the buffered output of the frequency PI controller is made available at a front panel BNC so that the user can monitor the frequency deviation.

The circuit also contains manual switches to bypass the bandpass filter, the amplitude control and phase offset is so desired.  These are not shown in Fig. \ref{FullPLL}.

\subsection{Digital VCO}

The output of the frequency PI controller is also fed into the input of an ADC (Analog Devices AD7894\cite{analog}).  The ADC, the Microchip PIC microcontroller (PIC 18F45K22\cite{microchip}) and the DDS chip (Analog Devices AD9832\cite{analog}) form the equivalent of the frequency-modulated waveform generator in the discrete PLL discussed earlier.  The PIC communicates with the ADC, the DDS and a 8-digit 7-segment display (not shown in Fig. \ref{FullPLL}) via a SPI interface.  While we could have incorporated the PIC microcontroller into the PCB we designed, we found it more convenient and very cost-effective to buy the Ready-for-PIC board as well as the display from MikroElektronika.\cite{mikro}  In addition to providing convenient connectors for interfacing with the other components of our PLL, the Ready-for-PIC board also comes with a USB interface that can be used to program the microcontroller and communicate with it once programmed.  We used MikroElektronika's MikroPascal development environment to 
write and download the program to the microcontroller.

The PIC controller sets the DDS frequency by sending it two 16-bit digital words over the SPI interface.  These two words define the frequency as a 32-bit fraction of the master clock frequency of the DDS.  The master clock  sets the minimum time increment or step for the DDS output.  Since our desired output frequency is in the range of 30 kHz, we have chosen a 1 MHz crystal oscillator to provide the master clock, giving a time step of 1 $\mu$s so that we have approximately 30 points to define each time period of the output wave form.  The AD7894 is a 14-bit ADC with a bipolar input of $\pm10$ V, providing a nominal resolution of $\sim$1.2 mV.   
\begin{figure}[!]
\center{\includegraphics[width=8.5cm]{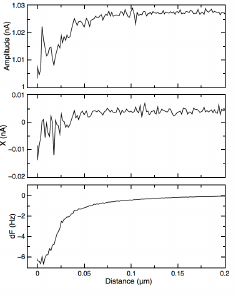}}
\caption{\small{Amplitude, $X$ output and frequency shift measured using the hybrid analog/digital PLL, as described in the text.}}
\label{PICPLL}
\end{figure}
The program that runs on the microcontroller is very simple.  On startup, it first initializes the 7-segment display and the ADC, and initializes and resets the DDS to output a sine wave at a default frequency, which we have chosen to be 32768 Hz (the nominal free resonant frequencies of our tuning forks).  It then goes into an infinite loop, where it monitors communications coming over the USB connection from the host computer.  These communications from the host computer are in ASCII, typically in the form of a capital letter followed by a number.  We have defined a limited set of 4 commands that are sufficient for specifying the operation of the device: 1)  `FREQ' followed by a number tells the PIC to output a sine wave at the specified center frequency; 2) `CONV' followed by a number specifies the multiplicative factor to convert the input read by the ADC into a frequency deviation, in mHz/V; 3) `BEGIN' by itself tells the PIC to set an internal flag (`PLL\_ON') specifying that it is in PLL mode; 4)  
`END' by itself tells the PIC to clear the PLL\_ON flag.  If PLL\_ON is set, the PIC then reads the voltage corresponding to the frequency deviation from the ADC, calculates the required frequency deviation based on the conversion factor, adds this frequency deviation to the center frequency, programs the DDS to output a sine wave at the new frequency, and updates the display to show the new frequency.  The loop is then repeated.  To control the instrument from the host computer, we have written a simple program in Free Pascal\cite{freepascal} using the Lazarus Integrated Development Environment.\cite{lazarus}  This control program can be readily integrated into our open source real-time scanning probe control software, RTSPM.\cite{chandra} 

\subsection{Performance of the hybrid analog/digital PLL}

Figure  \ref{PICPLL} shows the amplitude of the signal, $X$ output and frequency deviation in a close approach curve taken with the hybrid analog/digital PLL with conditions identical to Fig. \ref{PseudoPLL}, without engaging the amplitude feedback.  Unlike Fig. \ref{PseudoPLL}, the $X$ output corresponding to the phase of the oscillation does not change appreciably as a function of distance from the surface, showing that the frequency being tracked is truly the resonant frequency of oscillation of the tuning fork.  As a consequence, the frequency shift is almost 20 times as large as in Fig. \ref{PseudoPLL}.  This demonstrates the clear advantage of using a PI controller as the loop filter.
\begin{figure}[!]
\center{\includegraphics[width=8.5cm]{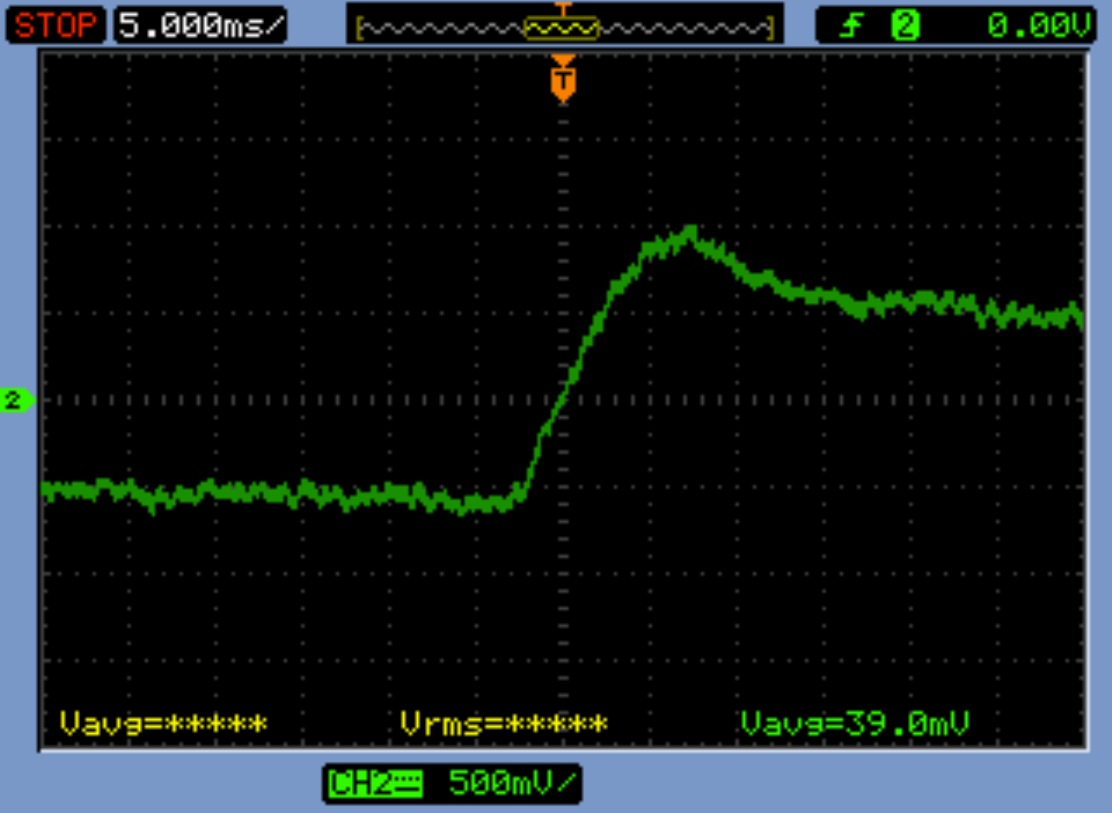}}
\caption{\small{Frequency deviation response of the hybrid analog/digital PLL to a step change in input frequency of 1 Hz.  The conversion factor is 2 Hz/V. }}
\label{TimeResponse}
\end{figure}

To demonstrate the time response of the PLL, we use the Agilent 33500B waveform generator and modulate its frequency to provide an input to the PLL, measuring the frequency deviation of the PLL as a function of time.  Fig. \ref{TimeResponse} shows the resulting curve, with the Agilent sourcing a sine wave of rms amplitude 1 V at 32768 Hz.  The frequency of the Agilent is modulated internally with a square wave of amplitude 1 Hz and frequency 2 Hz, and the conversion factor of the hybrid analog/digital PLL is 2 Hz/V.  It can be seen that the frequency deviation has a response time of a few ms, which is sufficient for most scanning applications.

In summary, we have developed a hybrid analog/digital PLL that provides the resolution and flexibility necessary for controlling non-contact force transducers in scanning probe microscopy.

\begin{acknowledgements}
This research was conducted with support from the
US Department of Energy, Basic Energy Sciences, under
grant number DE-FG02-06ER46346.
\end{acknowledgements}


\begin{thebibliography}{text} 
\bibitem{kalinin}S.V. Kalinin and A. Gruverman, Eds., \textit{Scanning Probe Microscopy}, Springer (2007).
\bibitem{giessibl}See, for example, Franz J. Giessibl, Appl. Phys. Lett. \textbf{76}, 1470 (2000).
\bibitem{horowitz}Paul Horowitz and Winfield Hill, \textit{The Art of Electronics}, 2nd Edition, Cambridge University Press (1989).
\bibitem{ti}For example, the popular CMOS CD54/74HC/HCT4046A (\url{http://www.ti.com}).
\bibitem{dds} \textit{Fundamentals of Direct Digital Synthesis (DDS)}, Analog Devices MT-085 Tutorial \url{http://www.analog.com/static/imported-files/tutorials/MT-085.pdf}.
\bibitem{signalrecovery}\url{http:://www.signalrecovery.com}.
\bibitem{agilent}\url{http://www.agilent.com}.
\bibitem{par}Princeton Applied Research , NJ, USA.  Unfortunately, this instrument is no longer made.
\bibitem{kleppner}D. Kleppner and R.J. Kolenkow, \textit{An introduction to mechanics}, McGraw-Hill (1973).
\bibitem{chandra}V. Chandrasekhar and M.M. Mehta, Rev. Sci. Instrum. \textbf{84}, 013705 (2013).
\bibitem{rozhok}S. Rozhok and V. Chandrasekhar, Solid State Communications \textbf{121}, 683 (2002).
\bibitem{nanosurf}See, for example, the EasyPLL from Nanosurf \url{http:\\www.nanosurf.com}.
\bibitem{microchip}Microchip Technology Inc., \url{http:\\www.microchip.com}.
\bibitem{supplemental}The Supplemental Materials contain two circuit schematics; (i) for the phase detector and low pass filter section and (ii) for the direct digital synthesis section of the overall implementation.
\bibitem{analog}Analog Devices Inc., \url{http:\\www.analog.com}.
\bibitem{mikro}Mikroelektronika, \url{http:\\www.mikroe.com}.
\bibitem{freepascal}\textit{Free Pascal: Advanced open source Pascal compiler for Pascal and Object Pascal}, \url{http://www.freepascal.org}.
\bibitem{lazarus}\textit{Lazarus}, \url{http://www.lazarus.freepascal.org}.


\end{thebibliography}
\end{document}


\maketitle

\newpage

\vspace{-4cm}

\begin{figure}
 \begin{center}
 \includegraphics[width=18cm]{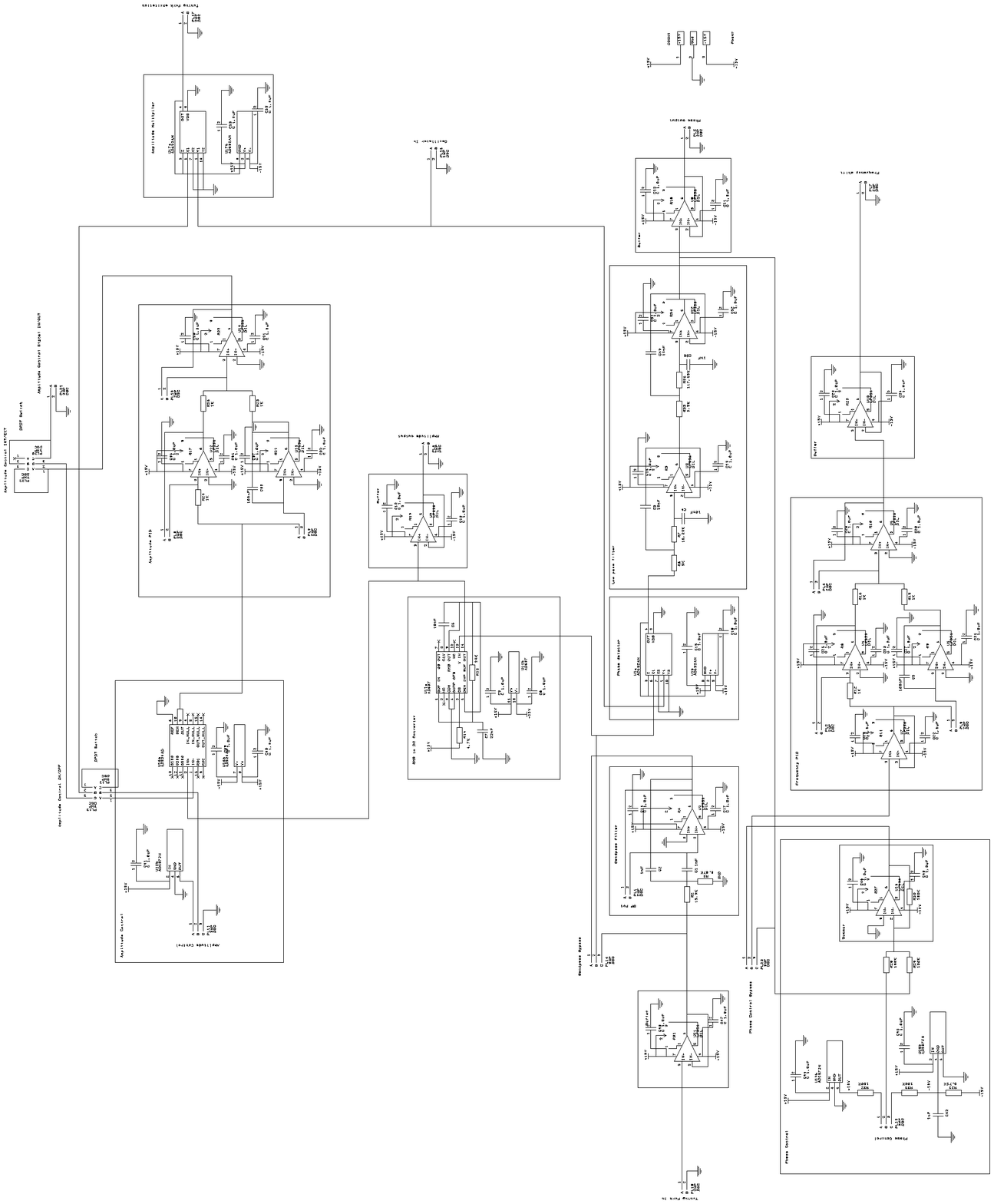}
 \end{center}
 \caption{Phase Detector Circuit}
\end{figure}

\newpage

\vspace{-4cm}

\begin{figure}
 \begin{center}
 \includegraphics[width=18cm]{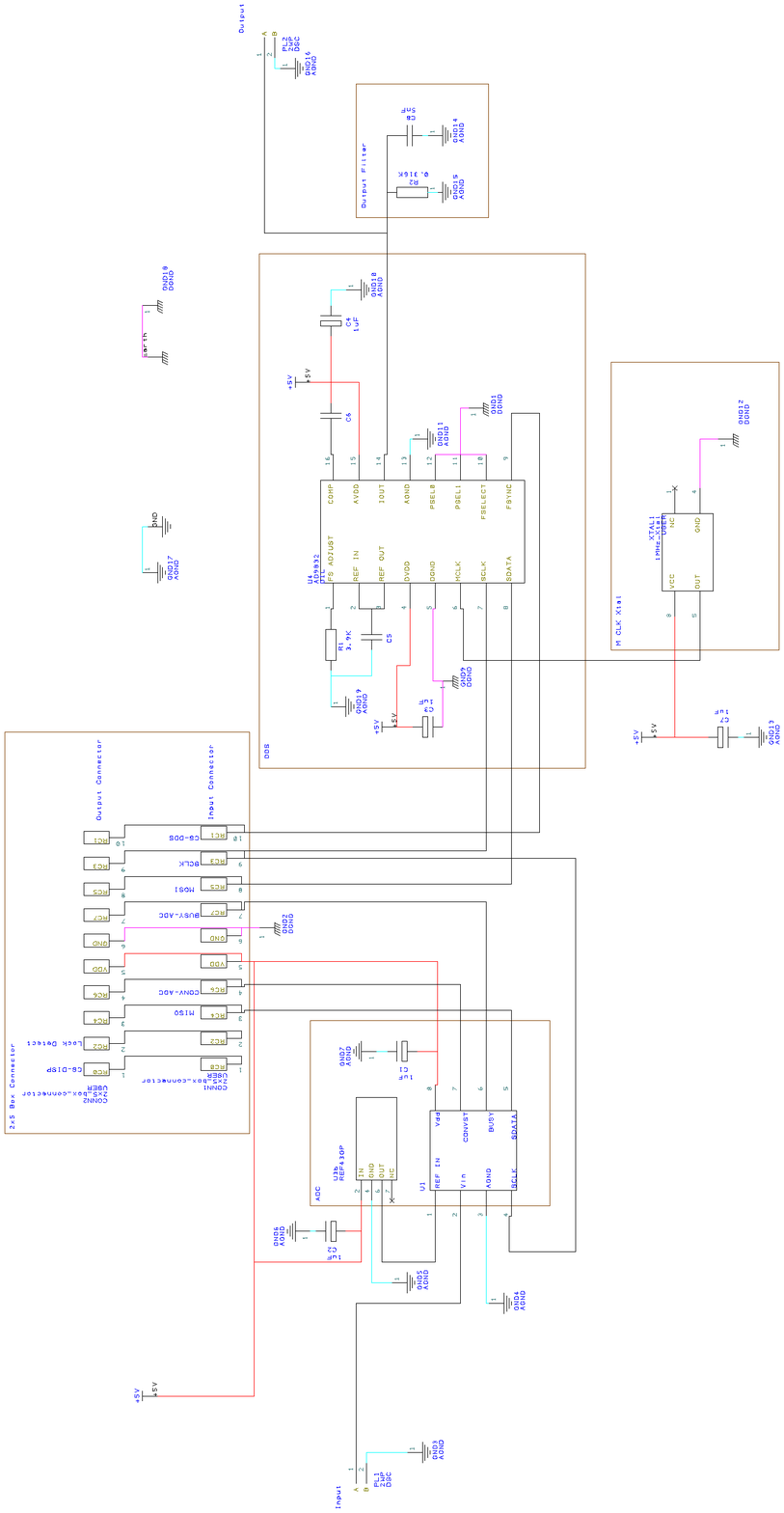}
 \end{center}
 \caption{Direct Digital Synthesis + ADC circuit}

 \end{figure}